\documentclass[aps,prl,twocolumn,superscriptaddress]{revtex4-2}
\usepackage{graphicx}
\usepackage{bm}
\usepackage{xcolor}
\usepackage{amsmath}
\usepackage{amssymb}
\usepackage{hyperref}
\usepackage{tabularray}

\begin{document}

\title{Spectral evidence for Dirac spinons in a kagome lattice antiferromagnet}

\author{Zhenyuan Zeng}
\affiliation{Beijing National Laboratory for Condensed Matter Physics, Institute of Physics, Chinese Academy of Sciences, Beijing 100190, China}
\affiliation{School of Physical Sciences, University of Chinese Academy of Sciences, Beijing 100190, China}
\author{Chengkang Zhou}
\affiliation{Department of Physics and HKU-UCAS Joint Institute of Theoretical and Computational Physics, The University of Hong Kong, Pokfulam Road, Hong Kong SAR, China}
\author{Honglin Zhou}
\affiliation{Beijing National Laboratory for Condensed Matter Physics, Institute of Physics, Chinese Academy of Sciences, Beijing 100190, China}
\affiliation{School of Physical Sciences, University of Chinese Academy of Sciences, Beijing 100190, China}
\author{Lankun Han}
\affiliation{Beijing National Laboratory for Condensed Matter Physics, Institute of Physics, Chinese Academy of Sciences, Beijing 100190, China}
\affiliation{School of Physical Sciences, University of Chinese Academy of Sciences, Beijing 100190, China}
\author{Runze Chi}
\affiliation{Beijing National Laboratory for Condensed Matter Physics, Institute of Physics, Chinese Academy of Sciences, Beijing 100190, China}
\affiliation{School of Physical Sciences, University of Chinese Academy of Sciences, Beijing 100190, China}
\author{Kuo Li}
\affiliation{Center for High Pressure Science and Technology Advanced Research, 10 Xibeiwang East Road, Haidian, Beijing 100094, China}
\author{Maiko Kofu}
\affiliation{J-PARC Center, Japan Atomic Energy Agency, Tokai, Ibaraki, 319-1195, Japan}
\author{Kenji Nakajima}
\email{kenji.nakajima@j-parc.jp}
\affiliation{J-PARC Center, Japan Atomic Energy Agency, Tokai, Ibaraki, 319-1195, Japan}
\author{Yuan Wei}
\affiliation{Photon Science Division, Paul Scherrer Institut, CH-5232 Villigen PSI, Switzerland}
\author{Wenliang Zhang}
\affiliation{Photon Science Division, Paul Scherrer Institut, CH-5232 Villigen PSI, Switzerland}
\author{Daniel G. Mazzone}
\affiliation{Laboratory for Neutron Scattering and Imaging, Paul Scherrer Institut, CH-5232 Villigen PSI, Switzerland}
\author{Zi Yang Meng}
\email{zymeng@hku.hk}
\affiliation{Department of Physics and HKU-UCAS Joint Institute of Theoretical and Computational Physics, The University of Hong Kong, Pokfulam Road, Hong Kong SAR, China}
\author{Shiliang Li}
\email{slli@iphy.ac.cn}
\affiliation{Beijing National Laboratory for Condensed Matter Physics, Institute of Physics, Chinese Academy of Sciences, Beijing 100190, China}
\affiliation{School of Physical Sciences, University of Chinese Academy of Sciences, Beijing 100190, China}
\affiliation{Songshan Lake Materials Laboratory, Dongguan, Guangdong, 523808, China}
\begin{abstract}
Emergent quasiparticles with a Dirac dispersion in condensed matter systems can be described by the Dirac equation for relativistic electrons, in analogy with Dirac particles in high-energy physics. For example, electrons with a Dirac dispersion have been intensively studied in electronic systems such as graphene and topological insulators. However, charge is not a prerequisite for Dirac fermions, and the emergence of Dirac fermions without charge degree of freedom has been theoretically predicted to be realized in Dirac quantum spin liquids. These quasiparticles carry a spin of 1/2 but are charge-neutral, and so are called spinons. Here we show that the spin excitations of a kagome antiferromagnet, YCu$_3$(OD)$_6$Br$_2$[Br$_{0.33}$(OD)$_{0.67}$], are conical with a spin continuum inside, which is consistent with the convolution of two Dirac spinons. The predictions of a Dirac spin liquid model with a spinon velocity obtained from the spectral measurements are in agreement with the low-temperature specific heat of the sample. Our results thus provide spectral evidence for the Dirac quantum spin liquid state emerging in this kagome lattice antiferromagnet. However, the locations of the conical spin excitations differ from those calculated by the nearest neighbor Heisenberg model, suggesting the Dirac spinons have an unexpected origin.
\end{abstract}

\maketitle

Quantum spin liquids (QSLs) provide an ideal platform for realizing quantum states of matter beyond the Landau paradigm of symmetry and its spontaneous breaking~\cite{BalentsL10,SavaryL17,ZhouY17,BroholmC20}. One of the crucial features of QSLs is the presence of fractionalized excitations, which have the form of elemental quasiparticles carrying the topological nature and interacting with the emergent gauge field~\cite{KivelsonSA87,WenXG91,WenXG17,GYSun2018,wangFractionalized2021,ChatterjeeA22}. Spinons are fractionalized excitations in the sense that they carry a spin of 1/2 but are charge-neutral. The spinons can be gapped or gapless. In the gapless state, the spinons can have a conical shape dispersion with the apex located at zero energy. This spectrum is like that of the Dirac cones in the electronic band structures of graphene and topological insulators \cite{GeimAK07,HasanMZ10,VafekO14}. Therefore, the spinons in Dirac QSLs are a new kind of Dirac quasiparticle without the charge degree of freedom. Although a Dirac QSL is interesting in its own right, it also serves as the parent state of novel two-dimensional quantum phases characterized by emergence and deconfinement~\cite{qinDuality2017,maDynamical2018,XYXu2019,SongXY19}. Although Dirac QSLs have been highly anticipated in many theoretical models of different kinds of lattices \cite{SachdevS92,HastingsMB00,RanY07,HermeleM08,IqbalY11,IqbalY14,HeYC17,LiaoHJ17,ZhuW19,ZhuZ18,LiuZX18,HuS19,SongXY20,NomuraY21,LiaoYD22}, their material realization has remained elusive due to the lack of spectral evidence.

\begin{figure}[tbp]
\includegraphics[width=\columnwidth]{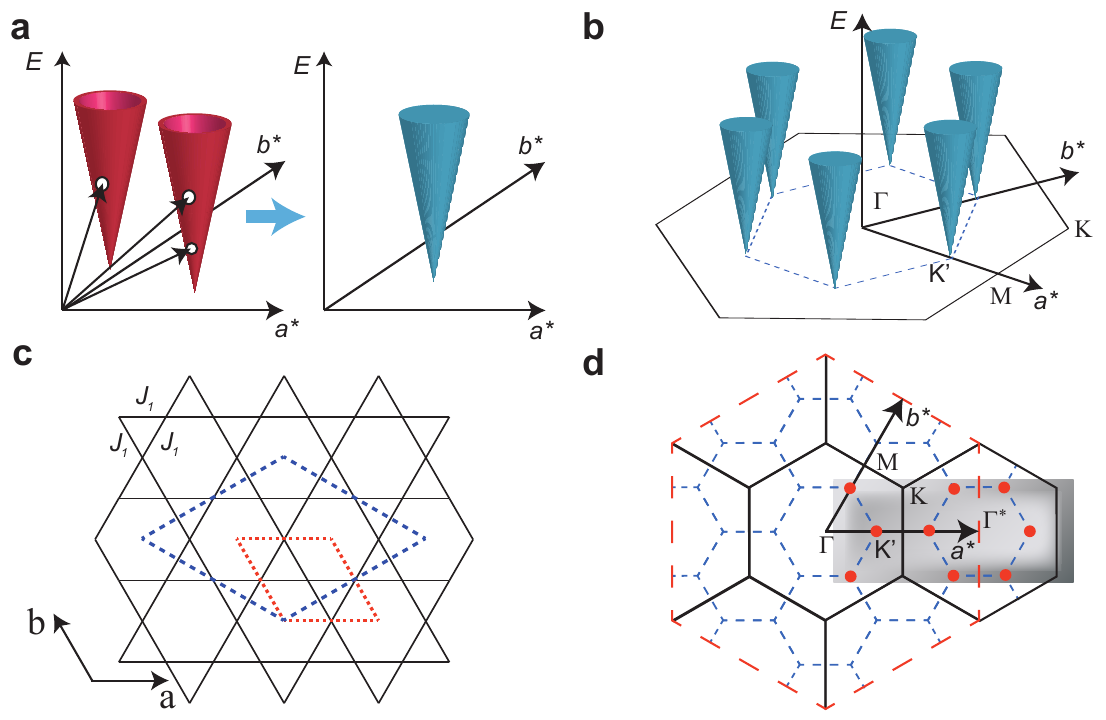}
 \caption{{\bf Schematics of the low-energy conical spin excitations and reciprocal space for YCu$_3$-Br.} {\bf a}, Schematic illustration of two Dirac spinons with a conical-surface dispersion (red) that merged into a cone spin excitations with continuum inside (blue). The two spinons can come from either two different Dirac cones or the same one as indicated by the white dots. {\bf b}, Six conical low-energy spin excitations in YCu$_3$-Br. Their momenta in the kagome Brillouin zone is indicated. {\bf c}, The in-plane kagome structure. The dashed red and blue lines represent the lattice unit cell of YCu$_3$-Br and Y$_3$Cu$_9$(OH)$_{19}$Cl$_8$ \cite{ChatterjeeD23}, respectively. {\bf d}, Sketch of the in-plane reciprocal space. The black solid line and the red and blue dashed lines represent the kagome, the extended kagome and the lattice of the Y$_3$Cu$_9$(OH)$_{19}$Cl$_8$ Brillouin zones, respectively. The gray shaded area illustrates the regime measured in this work. }
 \label{fig1}
\end{figure}

Although it is hard to directly detect single spinon excitations, two spinon excitations with total spin quantum number $S$ = 1 can result in a spin continuum that can be revealed by inelastic neutron scattering (INS). For Dirac spinons, one would expect that the spin excitations exhibit a convolved cone structure with a finite and continuous spectral weight inside the cone ( unlike the spin wave of linear dispersing magnons that have no weight inside the cone ) and with an apex at zero energy with respect to the ground state of the spin system \cite{NomuraY21,LiaoYD22}, as illustrated in Fig. \ref{fig1}a. In materials that have been theoretically suggested to host Dirac QSLs \cite{RanY07,ZhuZ18,HuS19}, the spin excitations observed in the INS experiments either do not look like Dirac spinons or are too blurry for their nature to be determined \cite{HanTH12,YaoS16,DingL19,BordelonMM20}. 

The discovery in this work changes the situation. The material studied in this work is YCu$_3$(OH)$_6$Br$_2$[Br$_{1-x}$(OH)$_{x}$] (denoted as YCu$_3$-Br hereafter). It has perfect kagome planes formed by Cu$^{2+}$ ions with $S$ = 1/2 \cite{ChenXH20}. Previous measurements have shown that there is no magnetic ordering down to 50 mK although its Weiss temperature is about -80 K \cite{ZengZ22,LiuJ22,LuF22,HongX22}. Moreover, the low-temperature specific heat shows a $T^2$ dependence at zero field and a $T$-linear term under a field \cite{ZengZ22}, which is consistent with a Dirac QSL \cite{RanY07}. Indeed, our neutron scattering results presented in this work for YCu$_3$-Br ($x$ = 0.67) clearly reveal six conical spin excitations with filled continuum weights inside the cones (Fig. \ref{fig1}b and \ref{fig1}d) . This represents strong evidence that the material realizes the long-sought-after Dirac QSL.

\begin{figure}[tbp]
\includegraphics[width=\columnwidth]{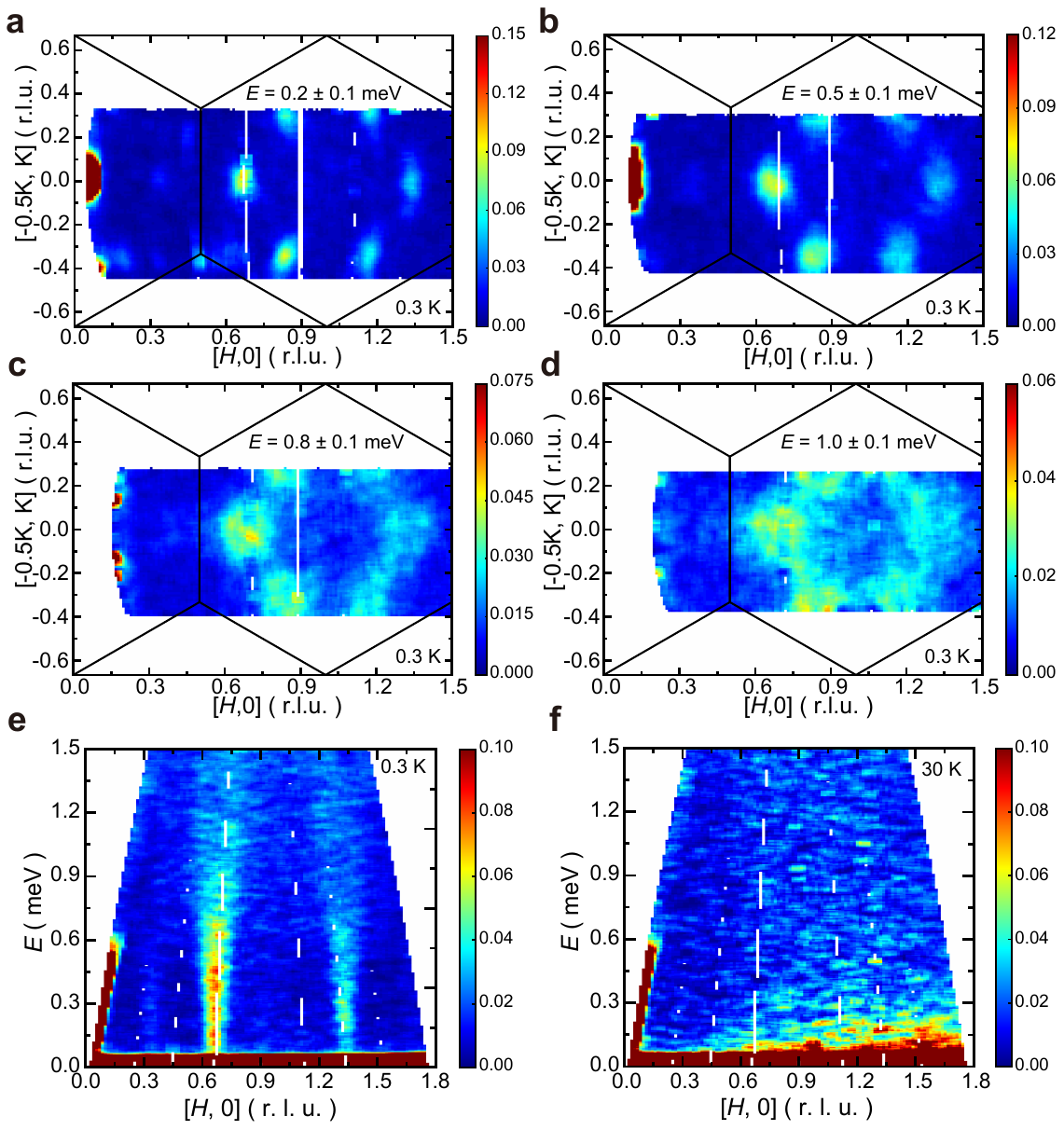}
 \caption{{\bf INS results at low energies with $E_i$ = 2.566 meV.} {\bf a-d}, Intensity contour plots of the INS results at 0.3 K in the [H,K] zone at 0.2 meV ({\bf a}), 0.5 meV ({\bf b}), 0.8 meV ({\bf c}) and 1 meV ({\bf d}). The $\pm$ sign gives the integrated energy range. The solid lines forming the hexagons mark the kagome Brillouin zones. The $x$ and $y$ axes are the components of $Q$ parallel to ($H$, 0) and (-0.5$K$, $K$), respectively.The positions of the excitations are at (1$\pm$1/3,0), (1,$\pm$1/3) and (1$\pm$1/3,$\mp$1/3). {\bf e}, {\bf f}, Intensity contour plots of the INS results as a function of $E$ and $Q$ along the [H,0] direction at 0.3 K ({\bf e}) and 30 K ({\bf f}). The integrated range along the [-$K$/2, $K$/2] is from $K$ = -0.05 to $K$ = 0.05. }
 \label{fig2}
\end{figure}

Figures \ref{fig2}a-\ref{fig2}d show contour plots of the in-plane low-energy excitations at several energies at 0.3 K for the deuterated YCu$_3$-Br. Six symmetrical spin excitations are centered at (1, 0) in the second Brillouin zone. The positions are at (2/3, 0) and the corresponding six-fold rotational points about the $c^*$-axis at (1, 0). Note that these positions correspond to the K' points in Fig. \ref{fig1}d. The peaks at 0.2 meV in Fig.~\ref{fig2}a are sharp and elongated along the direction vertical to $Q$ due to sample mosaic. With increasing energy (Fig.~\ref{fig2}b to Fig.~\ref{fig2}d), the peaks become broader, but no ring-like structure is observed at any energy and the excitations are always continuum with filled cones. Figure \ref{fig2}e shows an $E-Q$ plot with $Q$ along the [$H$, 0] direction. One can clearly see two filled cone-like excitations at (2/3, 0) and (4/3, 0), which disappear at 30 K (Fig. \ref{fig2}f), confirming the magnetic origin of these excitations. 

\begin{figure}[tbp]
\includegraphics[width=\columnwidth]{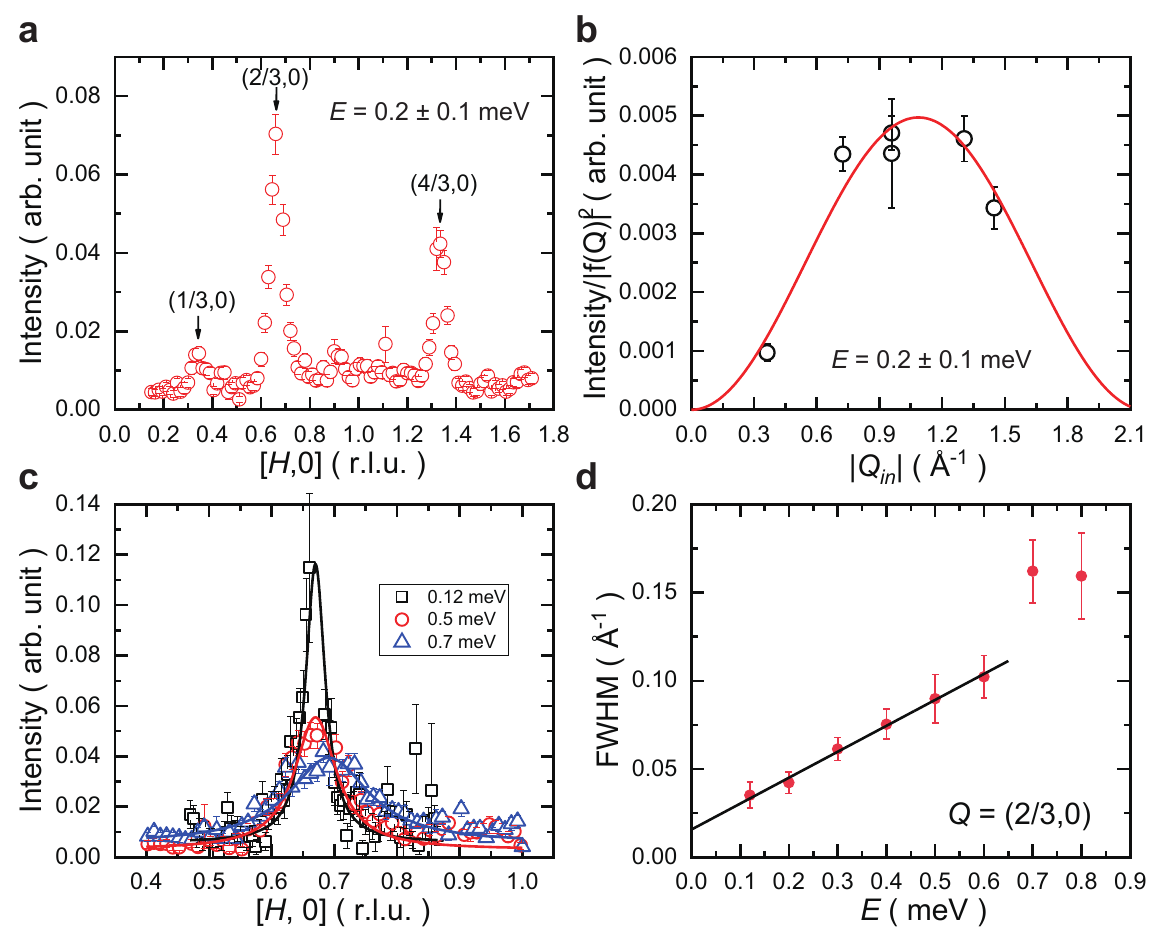}
 \caption{{\bf Quantitatively analysis for the low-energy data at 0.3 K.} {\bf a}, Constant-$E$ cuts along the [$H$, 0] direction at 0.2 meV. {\bf b}, The in-plane $|Q_{in}|$ dependence of the integrated intensity of the peaks at 0.2 meV, which has been normalized to the square of the magnetic form factor of the Cu$^{2+}$ ion. The solid line is the fit to a cosine function, as described in the main text. {\bf c}, Constant-$E$ cuts along the [$H$, 0] direction at several energies near (2/3, 0). The solid lines are fits to the Lorentzian function. {\bf d}, The energy dependence of the FWHM at $Q$ = (2/3, 0). The solid line is a linear fit. The error bars in all the panels represent one-standard-deviation uncertainty in the data based on Poisson statistics with the sample size being the neutron counts.}
 \label{fig3}
\end{figure}

To further quantitatively analyze the data, we made a cut along the [$H$, 0] direction at 0.2 meV, as shown in Fig. \ref{fig3}a. Besides the peaks at (2/3, 0) and (4/3, 0), there is also a peak at (1/3, 0). This suggests that the spin excitations also exist at the first Brillouin zone of the kagome lattice. Figure \ref{fig3}b shows the integrated intensity of the peaks normalized by the magnetic form factor as a function of the in-plane $|Q_{in}|$ = $|[H,K]|$. The integrated intensity is fitted well by $A(1-cos(\pi|Q_{in}|/|Q_{(1,0)}|))$, where $A$ is the only fitting parameter. This function explains why the intensity of the (1/3, 0) peak is much weaker and resembles the structure factor of randomly arranged nearest-neighbour singlets \cite{HanTH12}. In a similar system with long-range antiferromagnetic order \cite{ChatterjeeD23}, the spin excitations are also very strong in the second kagome Brillouin zone. Whether the fitting observed in Fig. \ref{fig3}b is coincidental or has a deeper physical meaning warrants further theoretical investigation. It is important to emphasize that the spin excitations around the peak positions do not follow the structure factor of randomly arranged nearest-neighbour singlets but exhibit a lorentzian behavior, as described below. Figure \ref{fig3}c shows the $Q$-cuts along the [H, 0] direction at (2/3, 0), which can all be well fitted by the Lorentzian function. The fitted full-width at half-maximum (FWHM) at 0.12 meV is about 0.035 $\pm$ 0.007 \AA$^{-1}$, which corresponds to a spin-spin correlation of about 180 \AA without considering the instrumental resolution. Moreover, the FWHM shows linear dependence below 0.6 meV (Fig.~\ref{fig3}d), which has a slope and intercept of 0.147 \AA$^{-1}$/meV and 0.016  \AA$^{-1}$, respectively. Note that the latter is very close to the instrumental resolution \cite{suppl}. The slope provides information about the velocity of spinons $\nu_F$ if the cone spin excitations come from the convolution of two Dirac spinons. The value is about 1$\times$10$^{3}$ m/s for free spinons and there could be renormalization due to interaction effects and the detailed fusion behaviour of two spinons. 

\begin{figure*}[tbp]
\includegraphics[width=\textwidth]{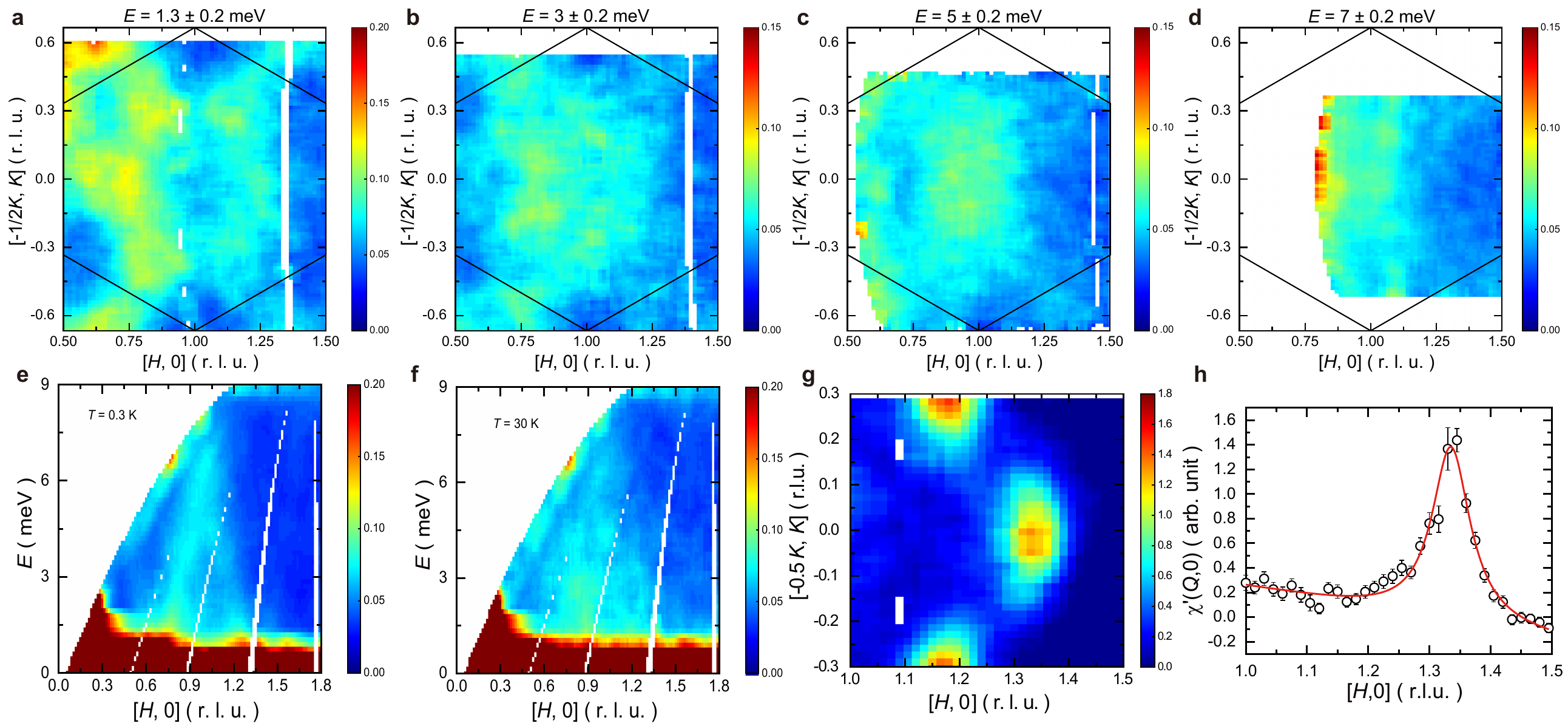}
 \caption{{\bf High-energy spin excitations at $E_i$ = 9.986 meV and static susceptibility $\chi'(Q,0)$ in YCu$_3$-Br.} {\bf a-d}, Intensity contour plots of the INS results at 0.3 K in the [H, K] zone at 1.3 meV ({\bf a}), 3 meV ({\bf b}), 5 meV ({\bf c}) and 7 meV ({\bf d}). {\bf e}, {\bf f}, Intensity contour plots of the INS results as a function of $E$ and $Q$ along the [H, 0] direction at 0.3 K ({\bf e}) and 30 K ({\bf f}).  Note that the intensities at small $Q$ near the edge of the colormap ({\bf c}-{\bf f}) are from the background. {\bf g}, $\chi'(Q,0)$ at 0.3 K calculated by the Kramers-Kronig relation based on $S(Q,\omega)$ between 0.075 to 8 meV. Only half of the Brillouin zone is shown to avoid the background at lower $Q$'s. {\bf h}, $H$ dependence of $\chi'(Q, 0)$ around (4/3, 0). The solid line is the fit to the Lorentzian function with a linear background. The error bars represent one-standard-deviation uncertainty in the data based on Poisson statistics with the sample size being the neutron counts.}
 \label{fig4}
\end{figure*}

The high-energy spin excitations of YCu$_3$-Br are shown in Fig. \ref{fig4}. At 1.3 meV, the excitations become so broad that they seem to be connected with those in adjacent zones (Fig. \ref{fig4}a). With increasing energy, the jointed six-branches spin excitations merge into one centered around (1, 0) and the area becomes smaller with increasing energy, as shown in Fig. \ref{fig4}b - \ref{fig4}d. The top of the excitations ends at about 7.5 meV (Fig. \ref{fig4}e), which is about 100 times the lowest energy observed. Unlike the low-energy spin excitations, the high-energy spin excitations survive at 30 K although there is substantial broadening (Fig. \ref{fig4}f).

As we can observe the whole spectra within the second kagome Brillouin zone, the static susceptibility $\chi'(Q, 0)$ can be obtained from the imaginary part of the dynamical susceptibility $\chi"(Q, \omega)$ by the Kramers-Kronig relation,
\begin{equation}
\chi'(Q,0) \propto \int_{-\infty}^{\infty}\frac{\chi"(Q,\omega)}{\omega}d\omega.
\label{eq1}
\end{equation}
\noindent According to the fluctuation-dissipation theorem, we have $\chi"(Q,\omega)$ = $S(Q,\omega)\times[1-\exp(-\hbar\omega/k_B T)]$, where $S(Q,\omega)$ is the scattering function that is proportional to the signal obtained from the INS measurements. Moreover, as $\chi"(Q,\omega)$ is the odd function of the energy, the integral can be done by just considering S(Q,$\omega$) at positive energy. The calculated result is shown in Fig. \ref{fig4}g, where there are still six symmetrical peaks. Figure \ref{fig4}h shows the $H$-cut at the (4/3, 0) peak and its Lorentzian fit. The magnetic correlation length $\xi$ is thus calculated to be about 83 \AA \ without considering the instrumental resolution, which is about 25 Cu-Cu bond lengths.

Our results provide strong evidence for an emerging Dirac QSL state in YCu$_3$-Br. The key finding in our work is the low-energy conical spin excitations with continuum inside, which requires a non-trivial origin of the low-energy excitations. For the conical spin excitations, trivial explanations such as damped spin waves in the presence of strong disorders cannot simultaneously reproduce the sharp excitations near zero energy and broad spectra at higher energies \cite{suppl}. The low-energy spinon continuum due to strong disorders always extend to a large area in the momentum space at low energies since disorder tends to destroy the spin-spin correlation and always reduces the spin-spin correlation length \cite{ZhuZ17,KimchiI18,MaZ21,ShimokawaT15,HanTH16}. In contrast, these features can be explained well by the presence of Dirac spinons, which results in conical spin excitations by two-spinon convolution \cite{RanY07}. The picture based on Dirac spinons also gives us a quantitative comparison between the INS and specific-heat measurements. According to a previous report \cite{ZengZ22}, the low-temperature specific heat of YCu$_3$-Br exhibits a quadratic temperature dependence at zero field, i.e., $C = \alpha T^2$. Theoretically, $\alpha$ can be calculated as $\alpha$ = 0.586 J/mol K$^3$ \cite{RanY07}, where we have used that $\nu_F \approx$ 10$^{3}$ m/s and there are six Dirac fermions. This is indeed very close to the experimental value, 0.452 J/mol K$^3$. Note that thermal-conductivity ($\kappa$) measurements on YCu$_3$-Br failed to detect a non-zero $\kappa/T|_{T\rightarrow 0 K}$ under magnetic fields \cite{HongX22}, which does not seem to agree with the existence of spinon Fermi surfaces induced by a field. However, as $\kappa$ = (1/3)$C\nu_Fl$, a very large mean free path $l$ ($>$ 30 $\mu$m) is required for the spinons to be detected by the heat-transport measurement ($\kappa/T$ $>$ 0.01 mW/K$^2$ cm). Such a large value of $l$ is apparently hard to achieve considering the site disorders in this system \cite{ChenXH20,ZengZ22,LiuJ22}. Also note that our results clearly show excitations down to 0.06 meV, which is much smaller than the gap value determined by the heat-transport measurement \cite{HongX22}, suggesting that the system has a gapless ground state.

The Dirac QSL state in YCu$_3$-Br is not the one predicted in the nearest Heisenberg model, which should exhibit conical spin excitations at (0.5, 0) (the point $M$ in Fig. \ref{fig1}d) \cite{ZhuW19}. The position of the spin excitations indicates that the magnetic system may be related to the $Q$ = $(1/3,1/3)$ phase \cite{HeringM22}, which suggests that there exist three different nearest exchange $J_1$ (Fig. \ref{fig1}c). Although this order seems closely related to Y$_3$Cu$_9$(OH)$_{19}$Cl$_8$ \cite{ChatterjeeD23}, which has distorted kagome planes and an enlarged lattice unit cell as shown in Fig. \ref{fig1}c, the kagome lattice remains undistorted in our sample. What is surprising is that the spin excitations close to zero energy are very sharp in our sample, and actually, much sharper than those in the ordered sample \cite{ChatterjeeD23}, which seems to suggest that the emergent Dirac spinons are insensitive to disorder. A detailed discussion on the spectrum of spin wave calculated in the presence of disorder in the form of different $J_1$ and the contrast with respect to our INS results are given in the Supplementary Information~\cite{suppl}. Whether the melting of the above order can result in a Dirac QSL is an open question. We note that an 1/9 magnetization plateau and magnetic quantum oscillations have been observed in this system \cite{ZhengG23}, which were explained by a model assuming that a Dirac spinon is coupled with the emergent gauge field, consistent with our observation of conical spin excitations. 

Finding a QSL state in the YCu$_3$-Br system seemingly with disorder is surprising, as it is believed that disorder will typically destroy QSLs \cite{DeyS20}. In our case, the disorder is mainly site disorder of Y$^{3+}$, Br$^-$, and (OH)$^{-}$, which results in alternate bonds of the hexagons in the kagome \cite{LiuJ22}. Note that the undistorted hexagons may also be treated as disorders to the hexagons with alternate bonds, which are related to the $Q$=$(1/3,1/3)$ phase \cite{HeringM22}, depending which quantity of hexagons is larger. Our results show that disorder has negligible effects on the low-energy spin excitations. This is consistent with the observation of magnetic quantum oscillations in this system \cite{ZhengG23}, which typically requires a clean system. This means that the Dirac QSL is either robust against disorder \cite{YangK22,SeifertUFP23} or even induced by disorder \cite{SavaryL17b,SibilleR17}. At this stage, an unambiguous theoretical calculation with disorder and longer-range interactions of the dynamic spectrum of frustrated quantum spin models is difficult and subject to certainly approximations such as finite size, finite temperature, finite bond-dimension and so on \cite{ShermanNE23}. On the other hand, the conic spin excitations emerging from Dirac spinons will render the very low-energy spin excitations with a sharp energy-momentum boundary \cite{qinDuality2017,maDynamical2018,SongXY20,NomuraY21}, as observed here. Overall, the YCu$_3$-Br system is an interesting platform for further experimental and theoretical studies.

We thank Prof. Yi Zhou, Yuan Li, and Patrick Lee for the discussions. This work is supported by the National Key Research and Development Program of China (Grants No. 2022YFA1403400, No. 2021YFA1400401), the K. C. Wong Education Foundation (Grants No. GJTD-2020-01), the Strategic Priority Research Program (B) of the Chinese Academy of Sciences (Grants No. XDB33000000),  and the Research Grants Council (RGC) of Hong Kong Special Administrative Region of China (Project Nos.  17301721, AoE/P701/20, 17309822, C7037-22GF, A\_HKU703/22 and 17302223). Measurements on AMATERAS were performed based on the approved proposal No. 2022B0048. The measurement on CAMEA was carried out under the proposal number 20220991.
We thank HPC2021 system under the Information Technology Services and the Blackbody HPC system at the Department of Physics, University of Hong Kong, as well as the Beijng PARATERA Tech CO.,Ltd. (URL: ttps://cloud.paratera.com).


%

\end{document}